\documentclass[
    ,final            
  ]
  {aipproc}

\layoutstyle{6x9}
\usepackage{epsfig}
\usepackage{amssymb}
\usepackage{amsmath}
\usepackage{hyperref}

\newcommand{\half}{{\frac{1}{2}}}
\newcommand{\mbf}[1]{\mathbf{#1}}

\begin{document}

\title{Light-Front Holography \\ and Novel Effects in QCD}


\author{Stanley J. Brodsky}{
  address={SLAC National Accelerator Laboratory, 
Stanford University, Stanford, CA 94309, USA }
  }
  \author{Guy F. de T\'eramond}{
  address={Universidad de Costa Rica, San Jos\'e, Costa Rica}
  }

\begin{abstract}

The correspondence between theories in anti-de Sitter space and
conformal field theories in physical space-time leads to an
analytic, semiclassical model for strongly-coupled QCD.  
Light-front holography allows hadronic amplitudes in the
AdS fifth dimension to be mapped to frame-independent light-front
wavefunctions of hadrons in physical space-time, thus providing a
relativistic description of hadrons at the amplitude level. We
identify the AdS coordinate $z$ with an invariant light-front
coordinate $\zeta$ which separates the dynamics of quark and gluon binding from 
the kinematics of constituent spin and internal orbital angular momentum. The result is a single-variable
light-front Schr\"odinger equation for QCD which determines the eigenspectrum and the light-front wavefunctions of hadrons for general spin and orbital angular momentum. The mapping of
electromagnetic and gravitational form factors in AdS space to their
corresponding expressions in light-front theory confirms this
correspondence. Some novel features  of QCD are discussed, including
the consequences of confinement for quark and gluon condensates and
the behavior of the QCD coupling in the infrared. The distinction
between static structure functions such as the probability
distributions  computed from the square of the light-front
wavefunctions versus dynamical structure functions which include the
effects of rescattering is emphasized. A new method for computing
the hadronization of quark and gluon jets at the amplitude level, an event amplitude generator, is outlined.

\end{abstract}

\maketitle
\section{Introduction}

One of the most challenging problems in strong interaction dynamics is to understand the interactions and composition of hadrons in terms of the fundamental quark and gluon degrees of freedom of the QCD Lagrangian. Because of the strong-coupling of QCD in the infrared domain, it has been difficult to find analytic solutions for the wavefunctions of hadrons or to make precise predictions for hadronic properties outside of the perturbative regime.  Thus an important theoretical goal is to find an initial approximation
to bound-state problems in QCD which is analytically tractable and which can be systematically improved.
The AdS/CFT correspondence~\cite{Maldacena:1997re} between string
states in anti--de Sitter (AdS) space and conformal field theories in physical space-time, modified for color confinement,
has led to a semiclassical model for strongly-coupled QCD which provides analytical insights
into its inherently non-perturbative nature, including hadronic spectra, decay constants, and wavefunctions.

As we have recently shown~\cite{Brodsky:2006uqa,Brodsky:2007hb}, there is a remarkable correspondence between the AdS description of hadrons and the Hamiltonian formulation of QCD in physical space-time quantized on the light front.  A key feature is ``light-front holography''  which allows one to precisely map the $\rm{AdS}_5$ solutions $\Phi(z)$ for hadron amplitudes  in the fifth dimensional variable $z$ to light-front wavefunctions $\psi_{n/H}$ of hadrons in the light-front coordinate $\zeta$ in physical
space-time~\cite{Brodsky:2006uqa,Brodsky:2007hb}, thus providing a relativistic
description of hadrons in QCD at the amplitude level. For two particles $\zeta^2 = \mbf{b}^2_\perp x(1-x)$ which is conjugate to the invariant mass squared ${\cal M}^2$ of the pair and also their light-front kinetic energy
 ${\mbf{k}^2_\perp}/{x(1-x)}.$  
One can derive this connection by showing that one
obtains the identical holographic mapping using the matrix elements
of the electromagnetic current and the energy-momentum tensor~\cite{Brodsky:2008pf}; this gives
an important consistency test and verification of holographic
mapping from AdS to physical observables defined on the light front.
The mathematical consistency of light-front holography for both the electromagnetic and gravitational~\cite{Brodsky:2008pf} hadronic transition matrix elements demonstrates that the mapping between the AdS holographic  variable $z$ and the transverse light-front variable $\zeta,$ which is a function of the multi-dimensional coordinates of the partons in a given light-front Fock state $x_i, \mbf{b}_{\perp i}$ at fixed light-front time $\tau,$ is a general principle.

Our analysis follows from recent developments in light-front
QCD~\cite{Brodsky:2006uqa,Brodsky:2007hb,Brodsky:2008pf,Brodsky:2008pg,Brodsky:2003px,deTeramond:2005su} which have been inspired by the
AdS/CFT correspondence~\cite{Maldacena:1997re}
between string states in anti-de Sitter (AdS) space and conformal
field theories (CFT) in physical space-time. 
Conformal symmetry is broken in physical QCD by quantum effects and quark masses.
The application of AdS space
and conformal methods to QCD can be motivated from the
empirical evidence~\cite{Deur:2008rf} and theoretical 
arguments~\cite{Brodsky:2008be} that the QCD
coupling $\alpha_s(Q^2) $ has an infrared fixed point at small $Q^2.$
In particular, a new extraction of the effective strong coupling constant
$\alpha_s^{g_1}(Q^2)$
from CLAS spin structure function data in an extended $Q^2$ region
using the Bjorken sum rule $\Gamma_1^{p-n}(Q^2)$~\cite{Deur:2008rf},
indicates the lack of $Q^2$ dependence of $\alpha_s$ in the low $Q^2$ limit. 
One can understand this
physically: in a confining theory where gluons have an effective 
mass~\cite{Cornwall:1981zr} or the  quarks and gluons have 
maximal wavelength~\cite{Brodsky:2008be}, all vacuum polarization
corrections to the gluon self-energy decouple at long wavelength.  Thus an infrared fixed point appears to be a natural consequence of confinement.  Furthermore, if one considers a
semiclassical approximation to QCD with massless quarks and without
particle creation or absorption, then the resulting $\beta$ function
is zero, the coupling is constant, and the approximate theory is
scale and conformal invariant~\cite{Parisi:1972zy}. One can then use conformal symmetry as 
a {\it template}, systematically correcting for its nonzero $\beta$ function as well
as higher-twist effects~\cite{Brodsky:1985ve}.  In particular, we can use the mapping of the group of Poincare and conformal generators $SO(4,2)$ to the isometries of $AdS_5$ space.  The fact that gluons have maximum wavelengths
and hence minimum momenta provides an explanation for why multiple emission of
soft gluons with large couplings does not spoil the Appelquist-Politzer
argument for the narrowness of $J/\psi$ and $\Upsilon$; the explanation is that
such emission is kinematically forbidden~\cite{Brodsky:2008be}.

In this lecture we review some of the features of the  ``hard-wall''~\cite{Polchinski:2001tt}  and ``soft-wall''~\cite{Karch:2006pv} AdS/QCD models~\cite{Erlich:2005qh,Da Rold:2005zs} which provide an initial approximation to QCD.  In our approach
the holographic mapping is carried out in the  strongly coupled regime where QCD is almost conformal, but in contrast with
the AdS/QCD framework described in~\cite{Karch:2006pv, Erlich:2005qh, Da Rold:2005zs}  quark and gluon degrees of freedom are explicitly introduced in the gauge/gravity correspondence. Consequently, the identification of orbital angular momentum of the constituents becomes a key element in the description of the internal structure of hadrons using holographic
principles. To this end, we also will discuss a procedure based on the light-front Hamiltonian of QCD~\cite{deTeramond:2008ht} which in principle allows the model to be systematically improved. 

Some of the important features of light-front AdS/QCD include:

(1) Effective frame-independent single-particle Sch\"odinger and Dirac equations for meson and baryon wave equations in both $z$ and $\zeta. $ 
For example, the meson eigenvalue equation is 
$$\big[-{d^2\over dz^2} - {1-4 L^2\over 4 z^2}+U(z) \big]\phi(z) = {\cal M}^2\phi(z).$$
where the soft-wall potential has the form of a harmonic oscillator  
$ U(z) = \kappa^4 z^2 + 2 \kappa^2(L+S-1).$

(2) The mass spectra formula for mesons  at zero quark mass in the soft-wall model is 
$${\cal M}^2 = 4 \kappa^2 (n + L +S/2),$$
which agrees with conventional Regge phenomenology. As in the Nambu string model based on a rotating flux tube, the Regge slope is the same for both the principal quantum number $n$ and the orbital angular momentum $L$.  The AdS/QCD correspondence thus builds in a remarkable connection between the string mass $\mu$ in the string  theory underlying an effective gravity theory in the fifth dimension with the orbital angular momentum of hadrons in physical space-time.

(3) The pion is massless at zero quark mass in agreement with general arguments based on chiral symmetry.

(4) The predicted form factors for the pion and nucleons agree well with experiment.  
The nucleon LFWFs  have both $S$ and $P$ wave-components allowing one to compute both the Dirac and Pauli form factors.
In general the AdS/QCD form factors fall off in momentum transfer squared $q^2$ with a leading power  predicted by the dimensional counting rules and the leading twist of the hadron's dominant interpolating operator at short distances. 
The short-distance behavior of a hadronic state is
characterized by its twist  (dimension minus spin) 
$\tau = \Delta - \sigma$, where $\sigma$ is the sum over the constituent's spin
$\sigma = \sum_{i = 1}^n \sigma_i$. Twist is also equal to the number of partons $\tau = n$ if $L=0.$
Under conformal transformations the interpolating operators transform according to their twist, and consequently the AdS isometries map the twist scaling dimensions into the AdS 
modes~\cite{Brodsky:2003px}. Light-front holography thus provide a
simple semiclassical approximation to QCD which has both constituent counting
rule behavior~\cite{Brodsky:1973kr,Matveev:1973ra} at short distances and confinement at large 
distances~\cite{Brodsky:2008pg,Polchinski:2001tt}.

(5) The timelike form factors of hadrons exhibit poles in the $J^{PC}= 1^{--}$ vector meson channels, an analytic feature  which arises from  the dressed electromagnetic current in AdS/QCD~\cite{Brodsky:2007hb,Grigoryan:2007my}.

(6) The form of the nonperturbative pion distribution amplitude $ \phi_\pi(x)$ obtained from integrating the $ q \bar q$ valence LFWF $\psi(x, \mbf{k}_\perp)$  over $\mbf{k}_\perp$,
has a quite different $x$-behavior than the
asymptotic distribution amplitude predicted from the ERBL PQCD
evolution~\cite{Lepage:1979zb,Efremov:1979qk} of the pion distribution amplitude.
The AdS prediction
$ \phi_\pi(x)  = \sqrt{3}  f_\pi \sqrt{x(1-x)}$ has a broader distribution than expected from solving the ERBL evolution equation in perturbative QCD.
This observation appears to be consistent with the results of the Fermilab diffractive dijet 
experiment~\cite{Aitala:2000hb}, the moments obtained from lattice QCD~\cite{Brodsky:2008pg} and pion form factor data~\cite{Choi:2006ha}.

\section{Light-Front Holography}

One of the most important theoretical tools in atomic physics is the
Schr\"odinger equation, which describes the quantum-mechanical
structure of atomic systems at the amplitude level. Light-front
wavefunctions (LFWFs) play a similar role in quantum chromodynamics, 
providing a fundamental description of the structure and
internal dynamics of hadrons in terms of their constituent quarks
and gluons. The light-front wavefunctions of bound states in QCD are
relativistic generalizations of the Schr\"odinger wavefunctions of
atomic physics, but they are determined at fixed light-cone time
$\tau  = t +z/c$ -- the ``front form'' introduced by
Dirac~\cite{Dirac:1949cp} -- rather than at fixed ordinary time $t$.
A remarkable feature of LFWFs is the fact that they are frame
independent; i.e., the form of the LFWF is independent of the
hadron's total momentum $P^+ = P^0 + P^3$ and $\vec P_\perp.$

When a flash from a camera illuminates a scene, each object is illuminated along the light-front of the flash; i.e., at a given $\tau$.  Similarly, when a sample is illuminated by an x-ray source such as the Linac Coherent Light  Source, each element of the target is struck at a given $\tau.$  In contrast, setting the initial condition using conventional instant time $t$ requires simultaneous scattering of photons on each constituent. 
Thus it is natural to set boundary conditions at fixed $\tau$ and then evolve the system using the light-front Hamiltonian $P^- = P^0-P^3 = i {d/d \tau}.$  The invariant Hamiltonian $H_{LF} = P^+ P^- - P^2_\perp$ then has eigenvalues $\mathcal{M}^2$ where $\mathcal{M}$ is the physical mass.   Its eigenfunctions are the light-front eigenstates whose Fock state projections define the light-front wavefunctions.

Light-front quantization is the ideal framework to describe the
structure of hadrons in terms of their quark and gluon degrees of
freedom. The simple structure of the light-front vacuum allows an unambiguous
definition of the partonic content of a hadron. Given the LFWFs, one
can compute observables such as hadronic form factors and structure
functions, as well as the generalized parton distributions and
distribution amplitudes which underly hard exclusive reactions. The
constituent spin and orbital angular momentum properties of the
hadrons are also encoded in the LFWFs.

A key step in the analysis of an atomic system such as positronium
is the introduction of the spherical coordinates $r, \theta, \phi$
which  separates the dynamics of Coulomb binding from the
kinematical effects of the quantized orbital angular momentum $L$.
The essential dynamics of the atom is specified by the radial
Schr\"odinger equation whose eigensolutions $\psi_{n,L}(r)$
determine the bound-state wavefunction and eigenspectrum. In this
paper, we show that there is an analogous invariant
light-front coordinate $\zeta$ which allows one to separate the
essential dynamics of quark and gluon binding from the kinematical
physics of constituent spin and internal orbital angular momentum.
The result is a single-variable light-front Schr\"odinger equation for QCD
which determines the eigenspectrum and the light-front wavefunctions
of hadrons for general spin and orbital angular momentum.

Light-Front Holography can be derived by observing the correspondence between matrix elements obtained in AdS/CFT with the corresponding formula using the LF 
representation~\cite{Brodsky:2006uqa} .  The light-front electromagnetic form factor in impact 
space~\cite{Brodsky:2006uqa,Brodsky:2007hb,Soper:1976jc} can be written as a sum of overlap of light-front wave functions of the $j = 1,2, \cdots, n-1$ spectator
constituents:
\begin{equation} \label{eq:FFb} 
F(q^2) =  \sum_n  \prod_{j=1}^{n-1}\int d x_j d^2 \mbf{b}_{\perp j}  \sum_q e_q
\exp \! {\Bigl(i \mbf{q}_\perp \! \cdot \sum_{j=1}^{n-1} x_j \mbf{b}_{\perp j}\Bigr)} 
\left\vert \tilde \psi_n(x_j, \mbf{b}_{\perp j})\right\vert^2.
\end{equation}
The formula is exact if the sum is over all Fock states $n$.
For definiteness we shall consider a two-quark $\pi^+$  valence Fock state 
$\vert u \bar d\rangle$ with charges $e_u = \frac{2}{3}$ and $e_{\bar d} = \frac{1}{3}$.
For $n=2$, there are two terms which contribute to the $q$-sum in (\ref{eq:FFb}). 
Exchanging $x \leftrightarrow 1-x$ in the second integral  we find ($e_u + e_{\bar d}$ = 1)
\begin{eqnarray} \nonumber
 F_{\pi^+}(q^2)  &=& \!  \int_0^1 \! d x \int \! d^2 \mbf{b}_{\perp}  
 e^{i \mbf{q}_\perp \cdot  \mbf{b}_{\perp} (1-x)} 
\left\vert \tilde \psi_{u \bar d/ \pi}\! \left(x,  \mbf{b}_{\perp }\right)\right\vert^2 \\
\label{eq:PiFFb}
&=&  2 \pi \int_0^1 \! \frac{dx}{x(1-x)}  \int \zeta d \zeta\, 
J_0 \! \left(\! \zeta q \sqrt{\frac{1-x}{x}}\right) 
\left\vert\tilde\psi_{u \bar d/ \pi}\!(x,\zeta)\right\vert^2,
\end{eqnarray}
where $\zeta^2 =  x(1-x) \mathbf{b}_\perp^2$ and $F_\pi^+(q\!=\!0)=1$. 
Notice that by performing an identical calculation for the
$\pi^0$ meson the result is $F_{\pi^0}(q^2) = 0$ for any value of $q$, as expected
from $C$-charge conjugation invariance.

We now compare this result with the electromagnetic form-factor in AdS space~\cite{Polchinski:2002jw}:
\begin{equation} 
F(Q^2) = R^3 \int \frac{dz}{z^3} \, J(Q^2, z) \vert \Phi(z) \vert^2,
\label{eq:FFAdS}
\end{equation}
where $J(Q^2, z) = z Q K_1(z Q)$.
Using the integral representation of $J(Q^2,z)$
\begin{equation} \label{eq:intJ}
J(Q^2, z) = \int_0^1 \! dx \, J_0\negthinspace \left(\negthinspace\zeta Q
\sqrt{\frac{1-x}{x}}\right) ,
\end{equation} we can write the AdS electromagnetic form-factor as
\begin{equation} 
F(Q^2)  =    R^3 \! \int_0^1 \! dx  \! \int \frac{dz}{z^3} \, 
J_0\!\left(\!z Q\sqrt{\frac{1-x}{x}}\right) \left \vert\Phi(z) \right\vert^2 .
\label{eq:AdSFx}
\end{equation}
Comparing with the light-front QCD  form factor (\ref{eq:PiFFb}) for arbitrary  values of $Q$
\begin{equation} \label{eq:Phipsi} 
\vert \tilde\psi(x,\zeta)\vert^2 = 
\frac{R^3}{2 \pi} \, x(1-x)
\frac{\vert \Phi(\zeta)\vert^2}{\zeta^4}, 
\end{equation}
where we identify the transverse light-front variable $\zeta$, $0 \leq \zeta \leq \Lambda_{\rm QCD}$,
with the holographic variable $z$.

Matrix elements of the energy-momentum tensor $\Theta^{\mu \nu} $ which define the gravitational form factors play an important role in hadron physics.  Since one can define $\Theta^{\mu \nu}$ for each parton, one can identify the momentum fraction and  contribution to the orbital angular momentum of each quark flavor and gluon of a hadron. For example, the spin-flip form factor $B(q^2)$, which is the analog of the Pauli form factor $F_2(Q^2)$ of a nucleon, provides a  measure of the orbital angular momentum carried by each quark and gluon constituent of a hadron at $q^2=0.$   Similarly,  the spin-conserving form factor $A(q^2)$, the analog of the Dirac form factor $F_1(q^2)$, allows one to measure the momentum  fractions carried by each constituent.
This is the underlying physics of Ji's sum rule~\cite{Ji:1996ek}:
$\langle J^z\rangle = \half [ A(0) + B(0)] $,  which has prompted much of the current interest in 
the generalized parton distributions (GPDs)  measured in deeply
virtual Compton scattering. Measurements of the GDP's are of particular relevance
for determining the distribution of partons in the transverse
impact plane, and thus could be confronted with AdS/QCD predictions which follow
from the mapping of AdS modes to the transverse impact representation~\cite{Brodsky:2006uqa}.
An important constraint is $B(0) = \sum_i B_i(0) = 0$;  i.e.  the anomalous gravitomagnetic moment of a hadron vanishes when summed over all the constituents $i$. This was originally derived from the equivalence principle of gravity~\cite{Teryaev:1999su}.  The explicit verification of these relations, Fock state by Fock state, can be obtained in the light-front quantization of QCD in  light-cone 
gauge~\cite{Brodsky:2000ii}.  Physically $B(0) =0$ corresponds to the fact that the sum of the $n$ orbital angular momenta $L$ in an $n$-parton Fock state must vanish since there are only $n-1$ independent orbital angular momenta.

The light-front expression for the helicity-conserving gravitational form factor in impact space
is~\cite{Brodsky:2008pf}
\begin{equation} \label{eq:Ab}
A(q^2) =  \sum_n  \prod_{j=1}^{n-1}\int d x_j d^2 \mbf{b}_{\perp j}  \sum_f x_f
\exp \! {\Bigl(i \mbf{q}_\perp \! \cdot \sum_{j=1}^{n-1} x_j \mbf{b}_{\perp j}\Bigr)} 
\left\vert \tilde \psi_n(x_j, \mbf{b}_{\perp j})\right\vert^2,
\end{equation}
which includes the contribution of each struck parton with longitudinal momentum $x_f$
and corresponds to a change of transverse momentum $x_j \mbf{q}$ for
each of the $j = 1, 2, \cdots, n-1$ spectators. 
For $n=2$, there are two terms which contribute to the $f$-sum in  (\ref{eq:Ab}). 
Exchanging $x \leftrightarrow 1-x$ in the second integral we find 
\begin{eqnarray} \label{eq:PiGFFb} \nonumber
A_{\pi}(q^2) &\! = \!&  2 \! \int_0^1 \! x \, d x \int \! d^2 \mbf{b}_{\perp}  
 e^{i \mbf{q}_\perp \cdot  \mbf{b}_{\perp} (1-x)} 
\left\vert \tilde \psi \left(x, \mbf{b}_{\perp }\right)\right\vert^2 \\
&\! = \!& 4 \pi \int_0^1 \frac{dx}{(1-x)}  \int \zeta d \zeta\,
J_0 \! \left(\! \zeta q \sqrt{\frac{1-x}{x}}\right)  \vert\tilde\psi(x,\zeta)\vert^2,
\end{eqnarray}
where $\zeta^2 =  x(1-x) \mathbf{b}_\perp^2$.  We now consider the expression for the hadronic gravitational form factor in AdS space~\cite{Abidin:2008ku}
\begin{equation} 
A_\pi(Q^2)  =  R^3 \! \! \int \frac{dz}{z^3} \, H(Q^2, z) \left\vert\Phi_\pi(z) \right\vert^2,
\end{equation}
where $H(Q^2, z) = \half  Q^2 z^2  K_2(z Q)$.
The hadronic form factor is normalized at $Q=0$, $A(0) = 1$.
Using the integral representation of $H(Q^2,z)$
\begin{equation} \label{eq:intHz}
H(Q^2, z) =  2  \int_0^1\!  x \, dx \, J_0\!\left(\!z Q\sqrt{\frac{1-x}{x}}\right) ,
\end{equation}
we can write the AdS gravitational form factor 
\begin{equation} 
A(Q^2)  =  2  R^3 \! \int_0^1 \! x \, dx  \! \int \frac{dz}{z^3} \, 
J_0\!\left(\!z Q\sqrt{\frac{1-x}{x}}\right) \left \vert\Phi(z) \right\vert^2 .
\label{eq:AdSAx}
\end{equation}
Comparing with the QCD  gravitational form factor (\ref{eq:PiGFFb}) we find an  identical  relation between the light-front wave function $\tilde\psi(x,\zeta)$ and the AdS wavefunction $\Phi(z)$
given in Eq. (\ref{eq:Phipsi}) which is obtained from the mapping of the pion electromagnetic transition amplitude.

\section{Light-Front Quantization of QCD}

One can  express the hadron four-momentum  generator $P =  (P^+, P^-,
\mbf{P}_{\!\perp})$, $P^\pm = P^0 \pm P^3$, in terms of the
dynamical fields, the Dirac field $\psi_+$, $\psi_\pm = \Lambda_\pm
\psi$, $\Lambda_\pm = \gamma^0 \gamma^\pm$, and the transverse field
$\mbf{A}_\perp$ in the $A^+ = 0$ gauge~\cite{Brodsky:1997de}
 \begin{eqnarray} \nonumber
P^-  &\!\!=\!&  \half  \! \int \! dx^- d^2 \mbf{x}_\perp  \, \bar \psi_+  \gamma^+
\frac{m^2 \!+ \left( i \mbf{\nabla}_{\! \perp} \right)^2}{ i \partial^+} \psi_+ + {\rm interactions}, \\
\nonumber
P^+ &\!\!=\!&   \int \! dx^- d^2 \mbf{x}_\perp \,
 \bar \psi_+ \gamma^+   i \partial^+ \psi_+, \\ \label{eq:P}
\mbf{P}_{\! \perp}  &\!\!=\!&  \half \int \! dx^- d^2 \mbf{x}_\perp \,
\bar \psi_+ \gamma^+   i \mbf{\nabla}_{\! \perp} \psi_+,
\end{eqnarray}
where the integrals are over the initial surface $x^+ = 0$, $x^\pm = x^0 \pm x^3$.
The operator $P^-$ generates LF time translations
$\left[\psi_+(x), P^-\right] = i \frac{\partial}{\partial x^+} \psi_+(x)$,
and the generators $P^+$ and $\mbf{P}_\perp$ are kinematical.
For simplicity we have omitted from (\ref{eq:P})
the contribution from the gluon field $\mbf{A}_\perp$.

The Dirac field operator is expanded as
\begin{equation} \label{eq:psiop}
\psi_+(x^- \!,\mbf{x}_\perp)_\alpha = \sum_\lambda \int_{q^+ > 0} \frac{d q^+}{\sqrt{ 2
 q^+}}
\frac{d^2 \mbf{q}_\perp}{ (2 \pi)^3} 
\left[b_\lambda (q)
u_\alpha(q,\lambda) e^{-i q \cdot x} + d_\lambda (q)^\dagger
v_\alpha(q,\lambda) e^{i q \cdot x}\right],
\end{equation}
with $u$ and $v$ LF spinors~\cite{Lepage:1980fj}. Similar expansion follows for the
gluon field $\mbf{A}_\perp$.
Using  LF
commutation relations 
$\left\{b(q), b^\dagger(q')\right\}
= (2 \pi)^3 \,\delta (q^+ \! - {q'}^+)
\delta^{(2)} \! \left(\mbf{q}_\perp\! - \mbf{q}'_\perp\right)$,
we find
\begin{equation*} \label{eq:Pm}
P^- \! =  \sum_\lambda \! \int \!  \frac{dq^+ d^2 \mbf{q}_\perp}{(2 \pi)^3 }   \,
\frac{m^2 \! + \mbf{q}_\perp^2}{q^+}  \,
 b_\lambda^\dagger(q) b_\lambda(q) + { \rm interactions},
\end{equation*}
and we recover the LF dispersion relation $q^- = ({\mbf{q}_\perp^2 + m^2})/{q^+}$, which follows
from the on shell relation $q^2 = m^2$. The LF time evolution operator
$P^-$ is conveniently written as a term which represents the sum of the kinetic energy of all the partons plus a sum of all the interaction terms.

It is convenient to define a LF Lorentz invariant Hamiltonian
$H_{LF}= P_\mu P^\mu = P^-P^+  \! - \mbf{P}^2_\perp$ with eigenstates
$\vert \psi_H(P^+, \mbf{P}_\perp, S_z )\rangle$
and eigenmass  $\mathcal{M}_H^2$, the mass spectrum of the color-singlet states
of QCD~\cite{Brodsky:1997de}
\begin{equation} \label{eq:HLF}
H_{LF} \vert \psi_H\rangle = {\cal M}^2_H \vert \psi_H \rangle.
\end{equation}
A state $\vert \psi_H \rangle$ is an expansion 
in multi-particle Fock states
$\vert n \rangle $ of the free LF Hamiltonian:
~$\vert \psi_H \rangle = \sum_n \psi_{n/H} \vert n \rangle$, where
a one parton state is $\vert q \rangle = \sqrt{2 q^+} \,b^\dagger(q) \vert 0 \rangle$.
The Fock components $\psi_{n/H}(x_i, {\mathbf{k}_{\perp i}}, \lambda_i^z)$
are independent of  $P^+$ and $\mbf{P}_{\! \perp}$
and depend only on relative partonic coordinates:
the momentum fraction
 $x_i = k^+_i/P^+$, the transverse momentum  ${\mathbf{k}_{\perp i}}$ and spin
 component $\lambda_i^z$. Momentum conservation requires
 $\sum_{i=1}^n x_i = 1$ and
 $\sum_{i=1}^n \mathbf{k}_{\perp i}=0$.
The LFWFs $\psi_{n/H}$ provide a
{\it frame-independent } representation of a hadron which relates its quark
and gluon degrees of freedom to their asymptotic hadronic state.

We can compute $\mathcal{M}^2$ from the hadronic matrix element
\begin{equation}
\langle \psi_H(P') \vert H_{LF}\vert\psi_H(P) \rangle  = 
\mathcal{M}_H^2  \langle \psi_H(P' ) \vert\psi_H(P) \rangle,
\end{equation}
expanding the initial and final hadronic states in terms of its Fock components. The computation is much simplified in the 
frame $P = \big(P^+, M^2/P^+, \vec{0}_\perp \big)$ where $H_{LF} =  P^+ P^-$.
We find
 \begin{equation} \label{eq:M}
 \mathcal{M}_H^2  =  \sum_n  \! \int \! \big[d x_i\big]  \! \left[d^2 \mbf{k}_{\perp i}\right]   
 \sum_q \Big(\frac{m_q^2 + \mbf{k}_{\perp q}^2}{x_q} \Big) 
 \left\vert \psi_{n/H} (x_i, \mbf{k}_{\perp i}) \right \vert^2  + {\rm interactions} ,
 \end{equation}
plus similar terms for antiquarks and gluons ($m_g = 0)$. The integrals in (\ref{eq:M}) are over
the internal coordinates of the $n$ constituents for each Fock state with phase space normalization
\begin{equation}
\sum_n  \int \big[d x_i\big] \left[d^2 \mbf{k}_{\perp i}\right]
\,\left\vert \psi_{n/H}(x_i, \mbf{k}_{\perp i}) \right\vert^2 = 1.
\end{equation}
The LFWF $\psi_n(x_i, \mathbf{k}_{\perp i})$ can be expanded in terms of  $n-1$ independent
position coordinates $\mathbf{b}_{\perp j}$,  $j = 1,2,\dots,n-1$, so that ~$\sum_{i = 1}^n \mbf{b}_{\perp i} = 0$.  We can also express (\ref{eq:M})
in terms of the internal coordinates $\mbf{b}_{\perp j}$ with $\mbf{k}_\perp^2  \to
- \nabla_{\mbf{b}_\perp}^2$.
The normalization is defined by
\vspace{-4pt}
\begin{equation}  \label{eq:Normb}
\sum_n  \prod_{j=1}^{n-1} \int d x_j d^2 \mathbf{b}_{\perp j}
\vert \psi_{n/H}(x_j, \mathbf{b}_{\perp j})\vert^2 = 1.
\end{equation}

To simplify the discussion we will consider a two-parton hadronic bound state.  In the limit
$m_q \to 0$
\begin{eqnarray} \nonumber
\mathcal{M}^2  &\!\!=\!\!&  \int_0^1 \! d x \! \int \!  \frac{d^2 \mbf{k}_\perp}{16 \pi^3}   \,
  \frac{\mbf{k}_\perp^2}{x(1-x)}
 \left\vert \psi (x, \mbf{k}_\perp) \right \vert^2  + {\rm interactions} \\ \label{eq:Mb}
  &\!\!=\!\!& \int_0^1 \! \frac{d x}{x(1-x)} \int  \! d^2 \mbf{b}_\perp  \,
  \psi^*(x, \mbf{b}_\perp)
  \left( - \mbf{\nabla}_{ {\mbf{b}}_{\perp}}^2\right)
  \psi(x, \mbf{b}_\perp) +   {\rm interactions}.
 \end{eqnarray}

 It is clear from (\ref{eq:Mb}) that the functional dependence  for a given Fock state is
given in terms of the invariant mass
\begin{equation}
 \mathcal{M}_n^2  = \Big( \sum_{a=1}^n k_a^\mu\Big)^2 = \sum_a \frac{\mbf{k}_{\perp a}^2}{x_a}
 \to \frac{\mbf{k}_\perp^2}{x(1-x)} \,,
 \end{equation}
 the measure of the off-mass shell energy~ $\mathcal{M}^2 - \mathcal{M}_n^2$.
 Similarly in impact space the relevant variable for a two-parton state is  $\zeta^2= x(1-x)\mbf{b}_\perp^2$.
Thus, to first approximation  LF dynamics  depend only on the boost invariant variable
$\mathcal{M}_n$ or $\zeta,$
and hadronic properties are encoded in the hadronic mode $\phi(\zeta)$ from the relation
\begin{equation} \label{eq:psiphi}
\psi(x,\zeta) = \frac{\phi(\zeta)}{\sqrt{2 \pi \zeta}} f(x).
\end{equation}
 We choose the normalization of  the LF mode $\phi(\zeta) = \langle \zeta \vert \phi \rangle$
 \begin{equation}
 \langle\phi\vert\phi\rangle = \int \! d \zeta \,
 \vert \langle \zeta \vert \phi\rangle\vert^2 = 1,
 \end{equation}
 and thus the prefactor $f(x)$ is normalized according to
$ \int_0^1 \! \frac{d x}{x(1-x)}  \vert f(x) \vert^2 = 1$.  The mapping  of transition matrix elements
 for arbitrary values of the momentum transfer~\cite{Brodsky:2006uqa, Brodsky:2008pf} 
 gives $f(x) = \sqrt{x(1-x)}$.

We can write the Laplacian operator in (\ref{eq:Mb}) in circular cylindrical coordinates
$(\zeta, \varphi)$ with $ \zeta = \sqrt{x(1-x)} \vert \mbf{b}_\perp \vert$:
$\nabla^2 = \frac{1}{\zeta} \frac{d}{d\zeta} \left( \zeta \frac{d}{d\zeta} \right)
+ \frac{1}{\zeta^2} \frac{\partial^2}{\partial \varphi^2}$, and factor out the angular dependence of the
modes in terms of the $SO(2)$ Casimir representation $L^2$ of orbital angular momentum in the
transverse plane:
$\phi(\zeta, \varphi) \sim e^{\pm i L \varphi} \phi(\zeta)$. Using  (\ref{eq:psiphi}) we find~\cite{deTeramond:2008ht}
\begin{eqnarray} \nonumber
\mathcal{M}^2  &\!\!=\!\!& \int \! d\zeta \, \phi^*(\zeta) \sqrt{\zeta}
\left( -\frac{d^2}{d\zeta^2} -\frac{1}{\zeta} \frac{d}{d\zeta}
+ \frac{L^2}{\zeta^2}\right)
\frac{\phi(\zeta)}{\sqrt{\zeta}}  
+ \int \! d\zeta \, \phi^*(\zeta) U(\zeta) \phi(\zeta) \\ \nonumber
&\!\!=\!\!&
\int \! d\zeta \, \phi^*(\zeta)
\left( -\frac{d^2}{d\zeta^2}
- \frac{1 - 4L^2}{4\zeta^2} +U(\zeta)\right)
\phi(\zeta),
\end{eqnarray}
where all the complexity of the interaction terms in the QCD Lagrangian is summed up in the effective potential $U(\zeta)$.
The light-front eigenvalue equation $H_{LF} \vert \phi \rangle = \mathcal{M}^2 \vert \phi \rangle$
is thus a light-front wave equation for $\phi$
\begin{equation} \label{eq:QCDLFWE}
\left(-\frac{d^2}{d\zeta^2}
- \frac{1 - 4L^2}{4\zeta^2} + U(\zeta) \right)
\phi(\zeta) = \mathcal{M}^2 \phi(\zeta),
\end{equation}
an effective single-variable light-front Schr\"odinger equation which is
relativistic, covariant and analytically tractable. Using (\ref{eq:M}) one can readily
generalize the equations to allow for the kinetic energy of massive
quarks~\cite{Brodsky:2008pg}.

In the hard-wall model one has $U(z)=0$; confinement is introduced by requiring the wavefunction to vanish at $z=z_0 \equiv 1/\Lambda_{\rm QCD}.$
In the case of the soft-wall model,  the potential arises from a ``dilaton'' modification of the AdS metric; it  has the form of a harmonic oscillator  $ U(z) = \kappa^4 z^2 + 2 \kappa^2(L+S-1).$
The term $ - (1-4 L^2)/ 4 z^2 $ is a contribution to the LF kinetic energy induced by the change of variables to $\zeta$.

Individual hadron states can be identified by their interpolating operator at $z\to 0.$  For example, the pseudoscalar meson interpolating operator
$\mathcal{O}_{2+L}= \bar q \gamma_5 D_{\{\ell_1} \cdots D_{\ell_m\}} q$, 
written in terms of the symmetrized product of covariant
derivatives $D$ with total internal space-time orbital
momentum $L = \sum_{i=1}^m \ell_i$, is a twist-two, dimension $3 + L$ operator
with scaling behavior determined by its twist-dimension $ 2 + L$. Likewise
the vector-meson operator
$\mathcal{O}_{2+L}^\mu = \bar q \gamma^\mu D_{\{\ell_1} \cdots D_{\ell_m\}} q$
has scaling dimension $\Delta=2 + L$.  The scaling behavior of the scalar and vector AdS modes $\Phi(z) \sim z^\Delta$ at $z \to 0$  is precisely the scaling required to match the scaling dimension of the local pseudoscalar and vector-meson interpolating operators.

\begin{figure}[!]
    \includegraphics[width=10cm]{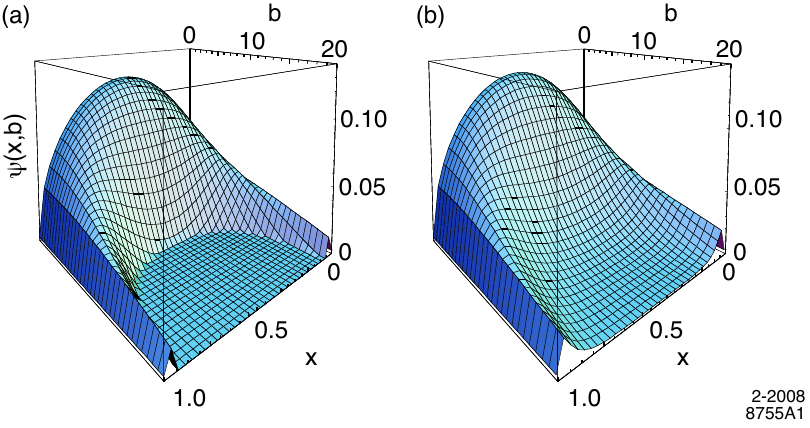}
  \caption{Pion light-front wavefunction $\psi_\pi(x, \mbf{b}_\perp$) for the  AdS/QCD (a) hard=wall ($\Lambda_{QCD} = 0.32$ GeV) and (b) soft-wall  ( $\kappa = 0.375$ GeV)  models.}
\label{fig2}  
\end{figure}

The resulting mass spectra  for mesons  at zero quark mass is
${\cal M}^2 = 4 \kappa^2 (n + L +S/2)$
in the soft-wall model.
The spectral predictions for both light meson and baryon states are compared with experimental data in~\cite{Brodsky:2008pg}.
The corresponding wavefunctions (see fig. \ref{fig2})
display confinement at large interquark
separation and conformal symmetry at short distances, reproducing dimensional counting rules for hard exclusive amplitudes.

\section {Vacuum Effects and Light-Front Quantization}

The light-front vacuum is remarkably simple in light-cone quantization because of the restriction $k^+ \ge 0.$   For example in QED,  vacuum graphs such as $e^+ e^- \gamma $  associated with the zero-point energy do not arise. In the Higgs theory, the usual Higgs vacuum expectation value is replaced with a $k^+=0$ zero mode~\cite{Srivastava:2002mw}; however, the resulting phenomenology is identical to the standard analysis.

Hadronic condensates play an important role in quantum chromodynamics (QCD).
Conventionally, these condensates are considered to be properties
of the QCD vacuum and hence to be constant throughout spacetime.
Recently  we have presented~\cite{Brodsky:2008be,Brodsky:2008xm,Brodsky:2008xu} a new perspective on the nature of QCD
condensates $\langle \bar q q \rangle$ and $\langle
G_{\mu\nu}G^{\mu\nu}\rangle$, particularly where they have spatial and temporal
support.
Their spatial support is restricted to the interior
of hadrons, since these condensates arise due to the interactions of quarks and
gluons which are confined within hadrons. For example, consider a meson consisting of a light quark $q$ bound to a heavy
antiquark, such as a $B$ meson.  One can analyze the propagation of the light
$q$ in the background field of the heavy $\bar b$ quark.  Solving the
Dyson-Schwinger equation for the light quark one obtains a nonzero dynamical
mass and, via the connection mentioned above, hence a nonzero value of the
condensate $\langle \bar q q \rangle$.  But this is not a true vacuum
expectation value; instead, it is the matrix element of the operator $\bar q q$
in the background field of the $\bar b$ quark.  The change in the (dynamical)
mass of the light quark in this bound state is somewhat reminiscent of the
energy shift of an electron in the Lamb shift, in that both are consequences of
the fermion being in a bound state rather than propagating freely.
Similarly, it is important to use the equations of motion for confined quarks
and gluon fields when analyzing current correlators in QCD, not free
propagators, as has often been done in traditional analyses of operator
products.  Since after a $q \bar q$ pair is created, the distance between the
quark and antiquark cannot get arbitrarily great, one cannot create a quark
condensate which has uniform extent throughout the universe.  The $55$ orders of magnitude conflict of QCD with the observed value of the cosmological condensate is thus removed~\cite{Brodsky:2008xu}.
A new perspective on the nature of quark and gluon condensates in
quantum chromodynamics is thus obtained:~\cite{Brodsky:2008be,Brodsky:2008xm,Brodsky:2008xu}  the spatial support of QCD condensates
is restricted to the interior of hadrons, since they arise due to the
interactions of confined quarks and gluons.  In the LF Theory, the condensate physics is replaced by the dynamics of higher non-valence Fock states. In particular, chiral symmetry is broken in a limited domain of size $1/ m_\pi$,  in analogy to the limited physical extent of superconductor phases.
This picture explains the
results of recent studies~\cite{Ioffe:2002be,Davier:2007ym,Davier:2008sk} which find no significant signal for the vacuum gluon
condensate.

\section{Hadronization at the Amplitude Level}

\begin{figure}[!]
\includegraphics[width=10cm]{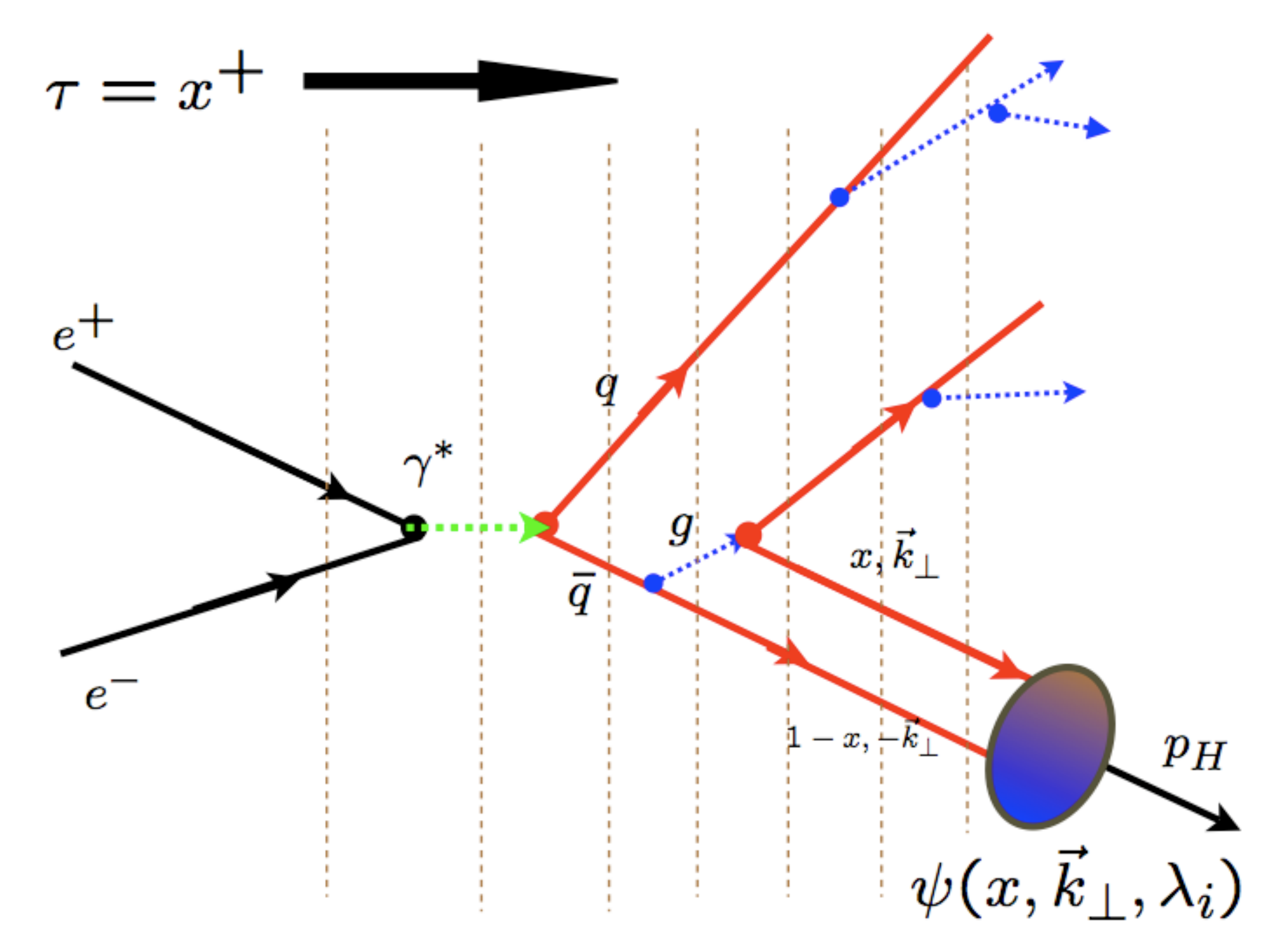}
  \caption{Illustration of an event amplitude generator for $e^+ e^- \to \gamma^* \to X$ for 
  hadronization processes at the amplitude level. Capture occurs if
  $\zeta^2 = x(1-x) \mbf{b}_\perp^2 > 1/ \Lambda_{\rm QCD}^2$
   in the AdS/QCD hard-wall model of confinement;  i.e., if
  $\mathcal{M}^2 = \frac{\mbf{k}_\perp^2}{x(1-x)} < \Lambda_{\rm QCD}^2$.}
\label{fig1}  
\end{figure}

The conversion of quark and gluon partons is usually discussed in terms  of on-shell hard-scattering cross sections convoluted with {\it ad hoc} probability distributions. 
The LF Hamiltonian formulation of quantum field theory provides a natural formalism to compute 
hadronization at the amplitude level~\cite{Brodsky:2008tk}.  In this case one uses light-front time-ordered perturbation theory for the QCD light-front Hamiltonian to generate the off-shell  quark and gluon T-matrix helicity amplitude  using the LF generalization of the Lippmann-Schwinger formalism:
\begin{equation}
T ^{LF}= 
{H^{LF}_I } + 
{H^{LF}_I }{1 \over {\cal M}^2_{\rm Initial} - {\cal M}^2_{\rm intermediate} + i \epsilon} {H^{LF}_I }  
+ \cdots 
\end{equation}
Here   ${\cal M}^2_{\rm intermediate}  = \sum^N_{i=1} {(\mbf{k}^2_{\perp i} + m^2_i )/x_i}$ is the invariant mass squared of the intermediate state and ${H^{LF}_I }$ is the set of interactions of the QCD LF Hamiltonian in the ghost-free light-cone gauge~\cite{Brodsky:1997de}.
The $T^{LF}$\!-matrix element is
evaluated between the out and in eigenstates of $H^{QCD}_{LF}$.   The event amplitude generator is illustrated for $e^+ e^- \to \gamma^* \to X$ in Fig. \ref{fig1}.

The LFWFS of AdS/QCD can be used as the interpolating amplitudes between the off-shell quark and gluons and the bound-state hadrons.
Specifically,
if at any stage a set of  color-singlet partons has  light-front kinetic energy 
$\sum_i {\mbf{k}^2_{\perp i}/ x_i} < \Lambda^2_{\rm QCD}$, then one coalesces the virtual partons into a hadron state using the AdS/QCD LFWFs.   This provides a specific scheme for determining the factorization scale which  matches perturbative and nonperturbative physics.

This scheme has a number of  important computational advantages:

(a) Since propagation in LF Hamiltonian theory only proceeds as $\tau$ increases, all particles  propagate as forward-moving partons with $k^+_i \ge 0$.  There are thus relatively few contributing
 $\tau-$ordered diagrams.

(b) The computer implementation can be highly efficient: an amplitude of order $g^n$ for a given process only needs to be computed once.  In fact, each non-interacting cluster within $T^{LF}$ has a numerator which is process independent; only the LF denominators depend on the context of the process.

(c) Each amplitude can be renormalized using the ``alternate denominator'' counterterm method~\cite{Brodsky:1973kb}, rendering all amplitudes UV finite.

(d) The renormalization scale in a given renormalization scheme  can be determined for each skeleton graph even if there are multiple physical scales.

(e) The $T^{LF}$\!-matrix computation allows for the effects of initial and final state interactions of the active and spectator partons. This allows for leading-twist phenomena such as diffractive DIS, the Sivers spin asymmetry and the breakdown of the PQCD Lam-Tung relation in Drell-Yan processes.

(f)  ERBL and DGLAP evolution are naturally incorporated, including the quenching of  DGLAP evolution  at large $x_i$ where the partons are far off-shell.

(g) Color confinement can be incorporated at every stage by limiting the maximum wavelength of the propagating quark and gluons.

(h) This method retains the quantum mechanical information in hadronic production amplitudes which underlie Bose-Einstein correlations and other aspects of the spin-statistics theorem.
Thus Einstein-Podolsky-Rosen QM correlations are maintained even between far-separated hadrons and  clusters.

A similar off-shell T-matrix approach was used to predict antihydrogen formation from virtual positron--antiproton states produced in $\bar p A$ 
collisions~\cite{Munger:1993kq}.

\section{Dynamical Effects of Rescattering}

Initial- and
final-state rescattering, neglected in the parton model, have a profound effect in QCD hard-scattering reactions,
predicting single-spin asymmetries~\cite{Brodsky:2002cx,Collins:2002kn}, diffractive deep lepton-hadron inelastic scattering~\cite{Brodsky:2002ue}, the breakdown of
the Lam Tung relation in Drell-Yan reactions~\cite{Boer:2002ju}, nor nuclear shadowing and non-universal 
antishadowing~\cite{Brodsky:2004qa}---leading-twist physics which is not incorporated in
the light-front wavefunctions of the target computed in isolation. 
It is thus important to distinguish~\cite{Brodsky:2008xe} ``static'' or ``stationary'' structure functions which are computed directly from the LFWFs of the target  from the ``dynamic'' empirical structure functions which take into account rescattering of the struck quark.   Since they derive from the LF eigenfunctions of the target hadron, the static structure functions have a probabilistic interpretation.  The wavefunction of a stable eigenstate is real; thus the static structure functions cannot describe diffractive deep inelastic scattering nor the single-spin asymmetries since such phenomena involves the complex phase structure of the $\gamma^* p $ amplitude.  
One can augment the light-front wavefunctions with a gauge link corresponding to an external field
created by the virtual photon $q \bar q$ pair
current~\cite{Belitsky:2002sm,Collins:2004nx}, but such a gauge link is
process dependent~\cite{Collins:2002kn}, so the resulting augmented
wavefunctions are not universal~\cite{Brodsky:2002ue,Belitsky:2002sm,Collins:2003fm}.  

\begin{figure}[!]
\includegraphics[width=13cm]{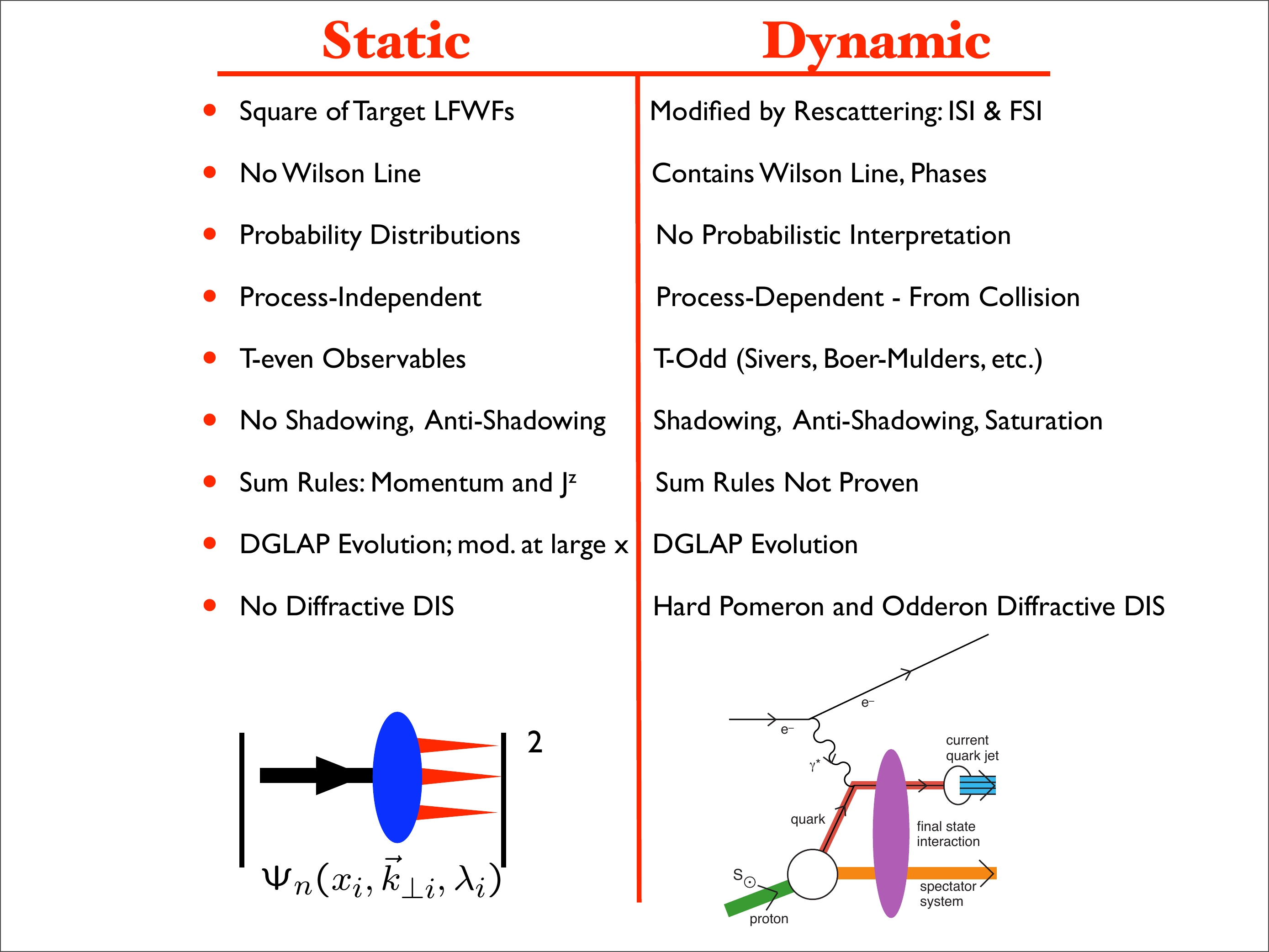}
\caption{Static versus dynamic structure functions}
\label{figstatdyn}  
\end{figure}

It should be emphasized
that the shadowing of nuclear structure functions is due to the
destructive interference between multi-nucleon amplitudes involving
diffractive DIS and on-shell intermediate states with a complex
phase.  The physics of rescattering and shadowing is thus not
included in the nuclear light-front wavefunctions, and a
probabilistic interpretation of the nuclear DIS cross section is
precluded. 
The distinction 
between static structure functions; i.e., the probability distributions  computed from the square of the light-front wavefunctions, versus the nonuniversal dynamic structure functions measured in deep inelastic scattering is summarized in fig. \ref{figstatdyn}.

\section{Conclusions}

Light-Front Holography is one of the most remarkable features of AdS/CFT.  It  allows one to project the functional dependence of the wavefunction $\Phi(z)$ computed  in the  AdS fifth dimension to the  hadronic frame-independent light-front wavefunction $\psi(x_i, \mbf{b}_{\perp i})$ in $3+1$ physical space-time. The 
variable $z $ maps  to $ \zeta(x_i, \mbf{b}_{\perp i})$. To prove this, we have shown that there exists a correspondence between the matrix elements of the electromagnetic current and the energy-momentum tensor of the fundamental hadronic constituents in QCD with the corresponding transition amplitudes describing the interaction of string modes in anti-de Sitter space with the external sources which propagate in the AdS interior. The agreement of the results for both electromagnetic and gravitational hadronic transition amplitudes provides an important consistency test and verification of holographic mapping from AdS to physical observables defined on the light-front.   The transverse coordinate $\zeta$ is closely related to the invariant mass squared  of the constituents in the LFWF  and its off-shellness  in  the light-front kinetic energy,  and it is thus the natural variable to characterize the hadronic wavefunction.  In fact $\zeta$ is the only variable to appear in the light-front
Schr\"odinger equations predicted from AdS/QCD.  

The use of the invariant coordinate $\zeta$ in light-front QCD
allows the separation of the dynamics of quark and gluon binding
from the kinematics of constituent spin and internal
orbital angular momentum. The result is a single-variable LF
Schr\"odinger equation  which determines the spectrum
and  LFWFs of hadrons for general spin and
orbital angular momentum. 
This LF wave equation serves as a first approximation to QCD and is equivalent to the
equations of motion which describe the propagation of spin-$J$ modes
on  AdS~\cite{deTeramond:2008ht}.
This allows us to establish
a gauge/gravity correspondence between an effective gravity theory
on AdS$_5$ and light front QCD.
The AdS/LF equations
correspond to the kinetic energy terms of  the partons inside a
hadron, whereas the interaction terms build confinement. Since
there are no interactions up to the confining scale in this approximation, there are no
anomalous dimensions. 
The eigenvalues of these equations for both meson and baryons give a good representation of the observed hadronic spectrum, especially in the case of the soft-wall model.  In the hard-wall model 
the orbital angular momentum decouples from the hadronic spin
$J$ and thus the LF excitation spectrum
of hadrons  depends only on 
orbital and principal quantum numbers. In the hard-wall 
model the dependence is linear:  $\mathcal{M} \sim 2n + L$. In the soft-wall
model the usual Regge behavior is found $\mathcal{M}^2 \sim n +
L$. Both models predict the same multiplicity of states for mesons
and baryons observed experimentally~\cite{Klempt:2007cp}. The predicted LFWFs have excellent phenomenological features, including predictions for the  electromagnetic form factors and decay constants. 
This may explain the experimental success of
power-law scaling in hard exclusive reactions where there are no
indications of  the effects of anomalous dimensions.

Nonzero quark masses are naturally incorporated into the AdS/LF predictions~\cite{Brodsky:2008pg} by including them explicitly in the LF kinetic energy  $\sum_i ( {\mbf{k}^2_{\perp i} + m_i^2})/{x_i}$. Given the nonpertubative LFWFs one can predict many interesting phenomenological quantities such as heavy quark decays, generalized parton distributions and parton structure functions.

We also note the distinction between between static structure functions such as the probability distributions  computed from the square of the light-front wavefunctions versus dynamical structure functions which include the effects of rescattering.  We have also shown that the LF Hamiltonian formulation of quantum field theory provides a natural formalism to compute hadronization at the amplitude level.

The AdS/QCD model is semiclassical, and thus it only predicts the lowest valence Fock state structure of the hadron LFWF. One can systematically improve the holographic approximation by
diagonalizing the QCD LF Hamiltonian on the AdS/QCD basis, or by
generalizing the variational and other systematic methods used in
chemistry and nuclear physics~\cite{Vary:2008bu}.
The action of the non-diagonal terms
in the QCD interaction Hamiltonian also generates the form of the higher
Fock state structure of hadronic LFWFs.  In
contrast with the original AdS/CFT correspondence, the large $N_C$
limit is not required to connect light-front QCD to
an effective dual gravity approximation.

\section*{Acknowledgments}
Presented by SJB at the XIII Mexican School of Particles and Fields,  San Carlos, Sonora, Mexico, October 2-11, 2008. He thanks Professors Alejandro Ayala, Maria Elena Tejeda-Yeomans, and their colleagues of the organizing committees for their outstanding hospitality in San Carlos.  We also thank Will Brooks, Carl Carlson, Stan Glazek, Paul Hoyer, Dae Sung Hwang, Pieter Maris, Ivan Schmidt, Robert Shrock and James Vary  for helpful conversations and collaborations. This research was supported by the Department
of Energy  contract DE--AC02--76SF00515.  SLAC-PUB-13491.

\end{document}